# AGS: An Dataset and Taxonomy for Domestic Scene Sound Event Recognition

Nan Che[1], Chenrui Liu[1], Fei Yu[2*], and Jiang Liu[1]

[1] Department of Computing Science and Technology,
Harbin University of Science and Technology, Harbin, 150040, China
[2] Zhejiang Lab, Hangzhou, 311121, China

**Abstract.** Environmental sound scene and sound event recognition is important for the recognition of suspicious events in indoor and outdoor environments (such as nurseries, smart homes, nursing homes, etc.) and is a fundamental task involved in many audio surveillance applications. In particular, there is no public common data set for the research field of sound event recognition for the data set of the indoor environmental sound scene. Therefore, this paper proposes a data set (called as **AGS**) for the home environment sound. This data set considers various types of overlapping audio in the scene, background noise. Moreover, based on the proposed data set, this paper compares and analyzes the advanced methods for sound event recognition, and then illustrates the reliability of the data set proposed in this paper, and studies the challenges raised by the new data set. Our proposed AGS and the source code of the corresponding baselines at https://github.com/taolunzu11/AGS .

## 1 Introduction

There is no existing sound dataset dedicated to domestic activities, although the sounds produced by domestic activities often have rich semantic features. Scene graph proposed by [1] focuses on the relationship between the entities to present semantics, it is naturally suitable for the expression of natural language [2]. The detailed semantics described in a scene graph can be a fundamental support to the tasks of activity recognition [3], image captioning [4], and so on. From then on, a large amount of research has paid attention to scene graph, and several visual scene graph datasets [1] [5] are released at the same time. Lots of scene graph generation (SGG) methods [6] [7] are studied in depth to improve the semantic expression quality of scene graph.

The data sets and research methods mentioned above are all based on the single modality of vision, acoustic modality which can be used to describe real-world scenarios in another aspect is usually ignored in the field of scene graph.

---

* The corresponding author: Fei Yu (yufei@zhejianglab.com)

The transmission of sound is naturally immune to line of sight which limits vision propagation, thus the sound scene can effectively fill the perception gap of the single visual scene, additionally, the characterization of sound in duration and intensity can provide a more in-depth description of activities which is hard in visual scenes. Finally, multi-modal recognition mixed with vision and sound may improve the performance of scene graph generation in the future.

Sound can also carry a lot of information worth extracting, several sound datasets are introduced to train classification models. Current sound data sets can be roughly divided into two categories: speech [8] [9] and nonspeech [10] [11], almost all sound datasets concentrate on sound classification, but cannot provide the relationship between entities that emit sounds. Therefore, we publish the Action Genome Sound Dataset (AGS) which extracts the sounds made by people and objects from the videos in the Action Genome dataset [12]. Similar to the original visual labels in Action Genome dataset, we manually label the attention relationship, spatial relationship, and contacting relationship between entities in the sound aspect.

Our AGS dataset contains 3986 sound records clipped from 972 video files and can be classified into a total of 65 categories, the number of relationship records reaches 4,260, additionally, the AGS dataset is mainly committed to domestic sounds to fill in the gap of current sound datasets. Then we conduct audio-only baseline experiments using six existing classification models such as MobileNetV2 [13], DaiNet [14], PANNs (Wavegram-Logmel-CNN) [15], Wavegram-Logmel-CNN-attention (our expanded method base on PANNs), AST [16] and LSTM-based methods[3], finally the accuracy and MAP of these models are presented.

In this paper, our main contributions are as the following:

- We extract sound records along the relationship dimensions of attention, space, and contacting, in addition to entities, classes, and clarity which are common labels in current sound datasets;
- We contribute a publicly available sound scene dataset that contains semantics by relationship description which is incremental work of AG dataset;
- We use early machine learning classifiers of audio to establish baseline audio classification performance on our dataset;
- The source code for baseline testing of AGS dataset are released in github.

## 2   Related Work

Audio (Sound) scene analysis (ASA) emphasizes the perception ability of human beings to perceive the environment and understand the environment through hearing, and ASA can provide technical and theoretical support for sound scene understanding and event recognition for scene monitoring. However, the current public and available data sets for sound scene understanding and sound event recognition cannot meet the research needs of sound scene understanding and sound event recognition. Compared with the related research on our work, it mainly includes **sound scene dataset** and **sound event recognition**.

---

[3] https://www.kaggle.com/code/kvpratama/audio-classification-with-lstm-and-torchaudio/notebook

## 2.1   Sound Scene Dataset

Most of the current public datasets mainly cover the following real-life scenes: transportation (cars, buses and trains), public spaces (grocery stores, restaurants, offices, streets and parks), and leisure spaces (beaches, basketball games and fields). Wherein the first sound (audio) scene dataset (Audioset) [17] mainly concluded the two millions clips by the manual annotation, which includes the main ontologies: human sounds, sounds of things (vehicle, engine, bell, alarm etc.), animal sounds, natural sound (wind,water, fire etc.), music. Audioset data does not distinguish between outdoor and indoor sound scenes. VGGSound [18] is a large-scale audiovisual dataset with low label noise collected from videos "in the wild", which is consisted of more than 200k videos for 300 audio classes. Both VGGSound and Audioset are based on youtube videos. FSD50K [19] fills the gap of AudioSet: AudioSet is not an open dataset because its official version contains precomputed audio features, which contianed over 51 k audio clips labeled using 200 classes. Different from the above YouTube data that does not distinguish between indoor and outdoor scenes, Mivia [20] for indoor scenes is synthetic, including 6K clips for 3 classes; DESED [21]for outdoor scenes is Freesound Dataset (FSD), including 12k clips for 10 classes; UrbanSound8k [22] for Audio Scenes is FSD, contains 8.7k for 10 classes.

## 2.2   Sound Event Recognition

The tasks of environmental audio scene recognition (EASR) and sound event recognition (SER) in uncontrolled environments are part of the computational auditory scene analysis (CASA) research field [23]. Sound event recognition (Sound event recognition) is to identify what and when is happening in an audio signal [24]. Most of the existing Sound event recognition methods are based on the above-mentioned public datasets, and none of the sound scenes they deal with can be directly applied to specific behavior monitoring. In the early days of Sound event recognition development, the existing research was mainly based on Mel-frequency cepstral coefficients (MFCCs) and machine learning models (SVM, HMM, GMM) [25−27]. Due to the wide application of deep learning in artificial intelligence, the SER approaches based on deep learning have gradually become the mainstream methods in ASA. Convolutional (CNN) models [28] have been widely used in AEDs. VGGish [28] is the first work based on CNN, which is also based on Audioset. The current representative methods mainly include MobileNetV2 [13], DaiNet [14], PANNs [15], VATT [29] and AST [16] etc. The core idea of MobileNetV2 is to take a low-dimensional compressed representation as input, first expand it to high-dimensional and use lightweight depthwise convolution for filtering. The features are subsequently projected back to a low-dimensional representation with linear convolution. DaiNet is different from previous studies that use log mel spectrogram as input, but directly uses time series waveform as input. PANNs is a Wavegram feature learnt from waveform, and a Wavegram-Logmel-CNN that achieves state-of-the-art performance in AudioSet tagging. In addition, transformer-based SER approaches (such as VATT and AST) [16,29] are mainly to use self-supervised learning to design the loss function and through the spectrogram features to integrate the vision transformer [30].

## 3    Domestic Scene Sound Taxonomy

### 3.1    The Specificity of AGS Record

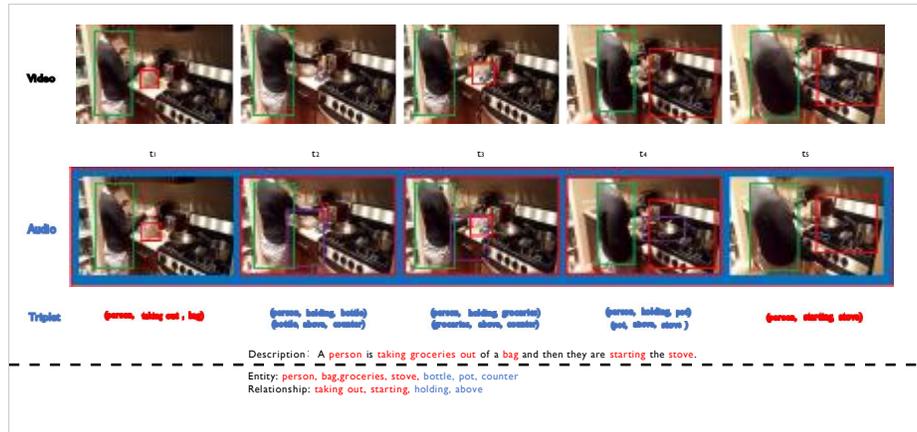

**Fig. 1.** An example of the typical advantage that auditory information has over visual information in environmental monitoring.

Shown as Fig. 1, a series of pictures are captured from a video named 0BXRP.mp4 in AG data, the red parts are the original record of the AG dataset, and the blue ones are the newly added entities, relationships, and triplets from the perspective of sound which are recorded in AGS dataset.

Specifically, entities in the video are labelled as person, bag, groceries, and stove in AG original data. Based on the combination of sound and video, we add new entities including bottle, pot, and counter to extend our AGS dataset. The triplets marked in red are the relationships in AG provided by the video description, while the blue triplets are the relationships annotated in AGS according to the sounds between entities. In summary, the AGS data set is an enhancement of the AG data in the sound relationship. The data application of AGS has more advantages in inference tasks, reasoning tasks, etc., because sound records

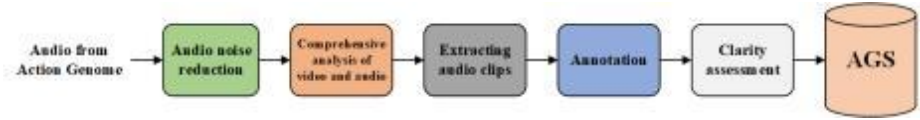

**Fig. 2.** The overall process of creating AGS.The process starts with the AG audio and AG Ontology.

can provide unique capabilities in the temporal dependence of events and the intensity of actions.

### 3.2 Recording Process and Overall Scale of AGS

The overall process of data labelling is shown in Fig. 2. Firstly, we select a video file from the AG dataset to prepare for audio processing. In the stage of audio noise reduction, we check whether the audio needs noise reduction, if necessary, noise reduction must process to filter the original background noise in the audio. In a comprehensive analysis of video and audio, we need to judge whether the clarity of this audio meets the requirements and whether the included soundscan correspond to the sound category and relationships, and if the audio is evaluated valueless, it will be discarded. In the stage of extracting audio clips, we extract several no more 4 seconds audio clips from the original audio that can reflect the sound category and relationship between entities. In the stage of annotation, We label the discovered entities and relationships. Finally, the records produced above are incrementally stored in the AGS dataset.

**Table 1.** Types of AGS Datasets and Description

| Types | Description | Samples |
|---|---|---|
| Clear Dataset | The data is clear enough, with almost no noise, to be able to clearly hear the category corresponding to the audio clip. | 1375 |
| Hybrid Dataset | The data is slightly noisy, but it identifies the category of audio clips. | 3986 |

After an amount of annotation work, there are currently 3,986 entries spanning 65 categories in our dataset. These 3,986 entries are clipped from 972 videos. Among them, 1,375 entries clipped from 682 video files are clear, which s, while 2,610 entries are unclear from 482 video files. We annotated 2,884 entries for human-object relationships and 1,376 entries for object-object relationships, with a total of 27 relationship categories.

**Table 2.** AG Classes VS AGS Classes

| classes | AG | AGS | classes | AG | AGS | classes | AG | AGS | classes | AG | AGS |
|---|---|---|---|---|---|---|---|---|---|---|---|
| curtain | | ✓ | phone__camera | ✓ | ✓ | sofa__couch | ✓ | ✓ | book | ✓ | ✓ |
| closestool | | ✓ | paper__notebook | ✓ | ✓ | shoe | ✓ | ✓ | blanket | | ✓ |
| stairs | | ✓ | mirror | ✓ | ✓ | refrigerator | ✓ | ✓ | bed | | ✓ |
| socket | | ✓ | medicine | | ✓ | pillow | ✓ | ✓ | bag | | ✓ |
| handset | | ✓ | light | ✓ | ✓ | picture | ✓ | ✓ | person | ✓ | ✓ |
| zipper | | ✓ | laptop | ✓ | ✓ | groceries | | ✓ | chopping board | | ✓ |
| spatula | | ✓ | food | ✓ | ✓ | laundry__detergent | | ✓ | knife | | ✓ |
| pot | | ✓ | floor | ✓ | ✓ | bowl | | ✓ | lipstick | | ✓ |
| dryer | | ✓ | doorknob | ✓ | ✓ | washing machine | | ✓ | sticky note | | ✓ |
| stove | | ✓ | door | ✓ | ✓ | clip | | ✓ | cap | | ✓ |
| pen | | ✓ | dish | ✓ | ✓ | pot cover | | ✓ | sprinkling can | | ✓ |
| water | ✓ | ✓ | cup__glass__bottle | ✓ | ✓ | trash can | | ✓ | hose | | ✓ |
| window | ✓ | ✓ | clothes | ✓ | ✓ | bell | | ✓ | circuit changer | | ✓ |
| vacuum | ✓ | ✓ | closet__cabinet | ✓ | ✓ | umbrella | | ✓ | ball | | ✓ |
| towel | ✓ | ✓ | chair | ✓ | ✓ | lighter | | ✓ | toys | | ✓ |
| television | ✓ | ✓ | broom | | ✓ | tape | ✓ | ✓ | sandwich | ✓ | |
| table | ✓ | ✓ | box | ✓ | ✓ | doorway | ✓ | | shelf | ✓ | |

As shown in Table 2, our dataset is annotated with 65 categories, while the AG dataset is annotated with 36 categories. It can be seen that our annotated categories cover most of the categories in the AG dataset. Among the 36 categories in the AG dataset, doorway, sandwich, and shelf are not referred in our sound dataset. However, we have set sandwich as a sub-label of food. In addition to these annotations, we have added 32 categories not present in the AG dataset. Our dataset is more detailed in its annotations, allowing richer information to be extracted from it.

**Table 3.** Classes of Clear Datasets and Description

| No. | Classes | # Samples | No. | Classes | # Samples |
|---|---|---|---|---|---|
| 1 | bag | 70 | 16 | laptop | 19 |
| 2 | bed | 12 | 17 | light | 41 |
| 3 | blanket | 6 | 18 | medicine | 23 |
| 4 | book | 37 | 19 | paper__notebook | 27 |
| 5 | box | 37 | 20 | person | 375 |
| 6 | broom | 28 | 21 | phone__camera | 7 |
| 7 | chair | 34 | 22 | picture | 3 |
| 8 | closet__cabinet | 61 | 23 | window | 10 |
| 9 | cup__glass__bottle | 100 | 24 | pillow | 9 |
| 10 | clothes | 25 | 25 | refrigerator | 33 |
| 11 | dish | 37 | 26 | shoe | 50 |
| 12 | door | 48 | 27 | sofa__couch | 6 |
| 13 | doorknob | 70 | 28 | television | 91 |
| 14 | floor | 15 | 29 | towel | 5 |
| 15 | food | 25 | 30 | vacuum | 71 |

As shown in Table 3, we present the sound categories and the number of samples corresponding to a particular category in the clear dataset as example.

## 4  Sound Event Recognition

### 4.1  Related Definition and Problem Descriptions

According to the reference [31], the formal definition of the sound event recognition (SER) is as follows:

**Definition 1. (Sound Event Recognition, SER):** Given a set of sound classes $C$, the subsets $Y = \bigcup_{c \in C} Y_c$ of ground truth labels for each sound class $c \in C$, wherein $Y_c = \{y_i = (t_{s,i}, t_{e,i}, c_i) : c_i = c\}$, $y_i = (t_{s,i}, t_{e,i}, c_i)$ represents each ground truth label $y_i$'s class $c_i$, the start time $t_{s,i}$ and the end time $t_{e,i}$, and given the subsets $X^* = \bigcup_{c \in C} X_c^*$ of the detections for each class $c \in C$, wherein $X_c^* = \{x_j = (t_{s,j}, t_{e,j}, c_j) : c_j = c\}$, $x_j = (t_{s,j}, t_{e,j}, c_j)$ , and where the starred notation $()^*$ indicates dependency on operating point parameters $\tau_c$. Then, the goal of the sound event recognition (SER) is to measure the performance of a system which output $X^*$ under the given information $Y$.

In the SER task, it is mainly necessary to recognize the label of the sound and the start and end time of the sound event. It can be seen that the accuracy of sound label recognition (also called as audio pattern recognition) plays an important role in the performance of SER. On the premise of accurate sound label recognition, sound events can be further guaranteed (in this paper, we represent sound events as a scene graph: nodes represent sound generation subject-object, edge represents the predicate relationship between subject-object) recognition accuracy (including the start and end time of the event).

### 4.2  SER-aware Baselines and Discussion

**SER-aware Baselines:** In order to further illustrate the contribution of the sound data of the home environment released in this paper for the sound pattern recognition task, we mainly conduct an experimental comparison analysis on the representative and current advanced algorithms in the current SER methods. Specifically, these approaches mainly include: MobileNetV2 [13], DaiNet [14], PANNs [15] and AST [16].

**Example:** In this section, we choose the sound event marker data with high sound clarity in the AGS data as the experimental data. We call such data **dataset of clear classes**[4], and these classes are shown in Table 3. In this part, we mainly use 1,406 sound clear data for experiments, which are extracted from 700 videos. In indoor sound environment recognition, due to the inherent sparsity of sound events. Therefore, we selected the top 6 categories from the 1,406 data for experiments,a total of 791 data, which were extracted from 439 videos.

---

[4] The dataset of clear classes: the data is clear enough, with almost no noise, to be able to clearly hear the category corresponding to the audio clip.

**Discussion:** We mainly use commonly used evaluation metrics in related research: Accuracy (Acc). The detailed experimental results areas shown on Table 4. In the process of running the example, the main hyperparameters are set as follows: the value range of the learning rate is:$\{1e^{-3}, 1e^{-4}, 1e^{-5}\}$; the ratio of the training set to the test set is: $\{5:5, 6:4, 7:3, 8:2\}$, and the iteration number is set to: $10,0000$.

**Table 4.** The results of Acc using different methods

| Train:Test | lr = $1e^{-3}$ | lr = $1e^{-4}$ | lr = $1e^{-5}$ | Train:Test | lr = $1e^{-3}$ | lr = $1e^{-4}$ | lr = $1e^{-5}$ |
|---|---|---|---|---|---|---|---|
| **PANNs** | | | | **DaiNet** | | | |
| 5:5 | 0.784 | 0.855* | 0.822 | 5:5 | 0.693 | 0.754* | 0.645 |
| 6:4 | 0.826 | 0.8237* | 0.829 | 6:4 | 0.722* | 0.709 | 0.636 |
| 7:3 | 0.869 | 0.92* | 0.899 | 7:3 | 0.772 | 0.810* | 0.781 |
| 8:2 | 0.867 | 0.892* | 0.867 | 8:2 | 0.722* | 0.722* | 0.646 |
| **MobileNetV2** | | | | **AST** | | | |
| 5:5 | 0.835* | 0.792 | 0.728 | 5:5 | 0.906 | 0.928* | 0.924 |
| 6:4 | 0.794* | 0.750 | 0.652 | 6:4 | 0.889 | 0.918* | 0.905 |
| 7:3 | 0.924* | 0.861 | 0.840 | 7:3 | 0.928 | 0.945* | 0.945* |
| 8:2 | 0.842* | 0.791 | 0.696 | 8:2 | 0.937 | 0.968* | 0.918 |

(* represents the best experimental results for each method under the same ratio of the training set and test set.)

From the experimental results in Table 4, it can be seen that the four methods have achieved good results on the AGS dataset. In particular, for the PANNs method, under the premise that the learning rate is set to lr = $1e^{-4}$ and the ratio of the training set to the test set is 7:3, the best experimental result is $0.92$. For the DaiNet method, under the premise that the learning rate is set to lr = $1e^{-4}$ and the ratio of the training set to the test set is 7:3, the best experimental result is $0.81$. For the MobileNetV2, under the premise that the learning rate is set to lr = $1e^{-3}$ and the ratio of the training set to the test set is 7:3, the best experimental result is $0.924$. For AST, under the premise that the learning rate is set to lr = $1e^{-4}$ and the ratio of the training set to the test set is 8:2, the best experimental result is $0.968$.

## 5   Experiment

### 5.1   Dataset Description

In the experimental part of this article, the amount of audio data is about $4,000$ pieces of data, which are extracted from $960+$ pieces of video. In this paper, the 20 categories with the largest number are selected from about 4000 pieces of data for experiments. From a total of 3300+ pieces of audio data set, these 3300+ pieces of audio data are extracted from 930+ pieces of video. The data set marked in this paper and the source code of the corresponding baseline method has been made public. The specific link is as follows: https://github.com/taolunzu11/AGS.

## 5.2   Experiment Setting

In our experiments, the main hyperparameters are set as follows: the value range of the learning rate is:$\{1e^{-3}, 1e^{-4}, 1e^{-5}\}$; the ratio of the training set to the test set is: $\{5 : 5, 6 : 4, 7 : 3, 8 : 2\}$, and the iteration number is set to: $\{10,000, 15,000, 20,000\}$. The computing environment is that Linux server with two NVIDIA Geforce RTX 3090, 125GB memory, and our experimental coding language is python. In addition, we mainly conduct an experimental comparison analysis on the representative and current advanced algorithms in the current SER methods. Specifically, these approaches mainly include: MobileNetV2 [13], DaiNet [14], PANNs (Wavegram-Logmel-CNN) [15], Wavegram-Logmel-CNN-attention (our expanded method base on PANNs), AST [16] and LSTM-based methods[5].

The brief description of these compared baseline methods follows:

- **MobileNetV2 [13]**: MobileNetV2 is to take a low-dimensional compressed representation as input, first expand it to high-dimensional and use lightweight depthwise convolution for filtering. The features are subsequently projected back to a low-dimensional representation with linear convolution;
- **DaiNet [14]**:  DaiNet is different from previous studies that use log mel spectrogram as input, but directly uses time series waveform as input;
- **AST [16]**: AST mainly utilizes the self-supervised learning to design the loss function and through the spectrogram features to integrate the vision transformer;
- **Wavegram-Logmel-CNN [15]**: Wavegram-Logmel-CNN is a Wavegram feature learnt from waveform, and combined with log mel spectrogram as input;
- **Wavegram-Logmel-CNN-attention**:In our work, we utilize the attention mechanism for the fusion methods of wavegram and log mel spectrogram in Wavegram-Logmel-CNN;
- **LSTM-based methods**: In this work, we use mel spectrogram and Mel-frequency cepstral coefficients extract audio features as input to the two-layer LSTM.

## 5.3   Evaluation Metrics

In order to further verify that the data set AGS proposed in this paper has the advantages of effective data support on the SER task and filling the shortage of existing Audioset data, we compare and analyze the performance of the current advanced SER methodson AGS through the evaluation metrics (MAP and Acc) recognized in related research.

---

[5] https://www.kaggle.com/code/kvpratama/audio-classification-with-lstm-and-torchaudio/notebook

– AP (Average Precision) [32]: The shape of the precision-recall curve at 11-point interpolation is summarized by averaging the maximum precision values at a set of equally spaced recall levels [0, 0.1, 0.2, ..., 1]. The formula is expressed as follows:

$$AP_{11} = \frac{1}{11} \sum_{R \in \{0,0.1, \ldots ,0.9,1\}} P_{interp}(R) \tag{1}$$

where

$$P_{interp}(R) = \max_{\tilde{R}:\tilde{R} \geq R} P(\tilde{R}) \tag{2}$$

To obtain the AP, consider the maximum precision $P_{interp}(R)$ for recall values greater than R in the above AP definition. In the all-point interpolation, interpolation is not only done using 11 equally spaced points but using all points available. The formula for full point interpolation is as follows:

$$AP_{all} = \sum_N (R_{n+1} - R_n) P_{interp}(R_{n+1}) \tag{3}$$

where

$$P_{interp}(R_{n+1}) = \max_{R:\tilde{R} \geq R_{n+1}} P(\tilde{R}) \tag{4}$$

In this AP definition, the precision at each level is obtained by interpolation, and the AP is obtained by taking the maximum precision at a recall value greater than or equal to $R_{n+1}$.

– MAP (Mean Average Precision) [32]: The mean AP (mAP) is a measure of the accuracy of the object detector for all classes in a particular database. The mAP is the average AP value for all categories.

$$mAP = \frac{1}{N} \sum_{i=1}^{N} AP_i \tag{5}$$

where $AP_i$ is the AP in class ith, and N is the total number of classes evaluated

– Acc (Accuracy) [33]: The proportion of the samples predicted correctly to the total sample, where the correct prediction may have a positive sample or anegative sample.

$$Accuracy(Acc) = \frac{TP + TN}{TP + TN + FP + FN} \tag{6}$$

with True positive (TP) is a correct detection of a ground-truth bounding box; False positive (FP)is an incorrect detection of a nonexistent object or a misplaced detection of an existing object; False negative (FN) is an undetected ground-truth bounding box; True negative (TN) is the correct rejection of a nonexistent objector a correct detection of the absence of an existing object.

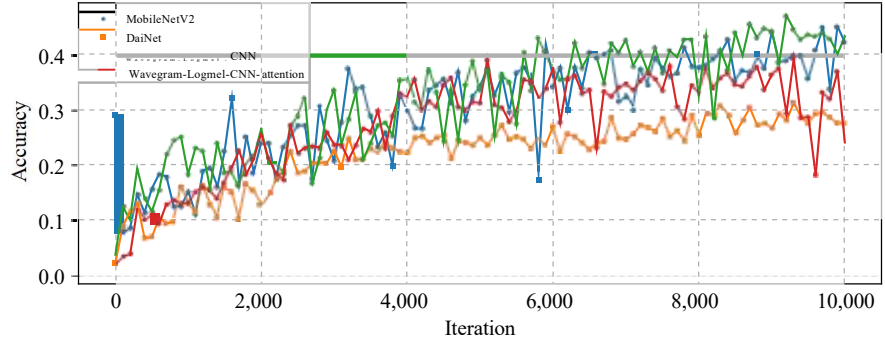

**Fig. 3.** Comparison of Sound Event Recognition for AGS using Acc

### 5.4    Experiment Results and Analysis

In this subsection, we compare 6 basleine methods. For the fairness of the experiment, all comparison methods are in the same experimental environment, and the corresponding hyperparameters are set to the best experimental results, while ensuring that the experimental results are in a state where both training and test results converge.

From the experimental results in Fig.3 and Fig.4, they can be seen that four baselines PANNs, PANNs-attention, DaiNet and MobileNetV2 have achieved better results on the AGS dataset. Since the training and testing process of these four methods are obtained through multiple iterations, however, both LSTM and AST are trained and tested through epoch, so here we mainly compare and analyze the experimental results of these four methods. The specific analysis is as follows:

**For the Accuracy (**Acc**) :**

- For the PANNs method, under the premise that the learning rate is set to lr = $1e^{-4}$ and the ratio of the training set to the test set is 8:2, the best experimental result (Acc) is 0.322 under the 10,000 iterations;

- For the PANNs-attention method, under the premise that the learning rate is set to lr = $1e^{-4}$ and the ratio of the training set to the test set is 8:2, the best experimental result (Acc) is 0.296 under the 10,000 iterations;

- For the DaiNet method, under the premise that the learning rate is set to lr = $1e^{-4}$ and the ratio of the training set to the test set is 8:2, the best experimental result (Acc) is 0.248 under the 15,000 iterations;

- For the MobileNetV2, under the premise that the learning rate is set to lr = $1e^{-3}$ and the ratio of the training set to the test set is 8:2, the best experimental result (Acc) is 0.307 under the 10,000 iterations.

**For the Mean Average Precision (**MAP**):**

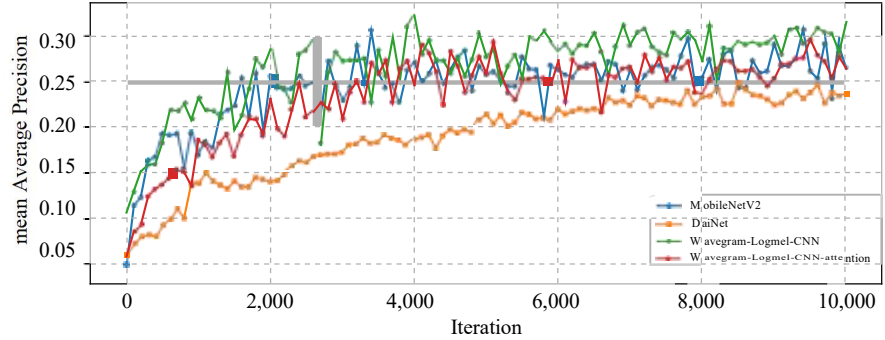

**Fig. 4.** Comparison of Sound Event Recognition for AGS using MAP

– For PANNs, under the premise that the learning rate is set to lr = 1e⁻⁴ and the ratio of the training set to the test set is 8:2, the best experimental result (MAP) is 0.473 under the 10,000 iterations;

– For PANNs-attention, under the premise that the learning rate is set to lr = 1e⁻⁴ and the ratio of the training set to the test set is 8:2, the best experimental result (MAP) is 0.395 under the 10,000 iterations;

– For DaiNet, under the premise that the learning rate is set to lr = 1e⁻⁴ and the ratio of the training set to the test set is 8:2, the best experimental result (MAP) is 0.35 under the 20,000 iterations;

– For the MobileNetV2, under the premise that the learning rate is set to lr = 1e⁻³ and the ratio of the training set to the test set is 8:2, the best experimental result (MAP) is 0.453 under the 10,000 iterations.

**Table 5.** Comparison of Six Baselines for AGS using Acc

| Method | lr | Train:Test | Acc | Iteration/Epoch | Comparison |
|---|---|---|---|---|---|
| MobileNetV2 | 1e⁻³ | 8:2 | 0.453 | 10,000 | ○ 3 |
| DaiNet | 1e⁻⁴ | 8:2 | 0.35 | 20,000 | ○ 5 |
| Wavegram-Logmel-CNN | 1e⁻⁴ | 8:2 | 0.473 | 10,000 | ○ 2 |
| Wavegram-Logmel-CNN-attention | 1e⁻⁴ | 8:2 | 0.395 | 10,000 | ○ 4 |
| LSTM-based Method | 1e⁻³ | 5:5 | 0.289 | 6 (epoch) | ○ 6 |
| AST | 1e⁻⁴ | 7:3 | 0.498 | 100 (epoch) | ○ 1 |

In order to compare and analyze all comparison methods on AGS, based on the best experimental settings (hyperparameter settings, experimental environment, loss function settings and optimization choices), the specific experimental results can be obtained in detail through Table 5 and Table 6.

**Table 6.** Comparison of Six Baselines for AGS using MAP

| Method | lr | Train:Test | MAP | Iteration/Epoch | Comparison | |
|--------|-----|-----------|------|----------------|-----------|---|
| MobileNetV2 | $1e^{-3}$ | 8:2 | 0.307 | 10,000 | ○3 | — |
| DaiNet | $1e^{-4}$ | 8:2 | 0.248 | 15,000 | ○5 | — |
| Wavegram-Logmel-CNN | $1e^{-4}$ | 8:2 | 0.322 | 10,000 | ○2 | — |
| Wavegram-Logmel-CNN-attention | $1e^{-4}$ | 8:2 | 0.296 | 10,000 | ○4 | — |
| LSTM-based Method | $1e^{-4}$ | 5:5 | 0.121 | 6 (epoch) | ○6 | — |
| AST | $1e^{-4}$ | 7:3 | 0.348 | 100 (epoch) | ○1 | — |

(— indicates that there is no change in ranking compared to the experimental results in Table 5.)

It can be seen from Table 5 that the evaluation indicators Acc of the six methods are all above 30% except for the LSTM-based method. According to the experimental results of SER on Audioset in related research [11] (The Acc result of the better method is 30% to 35%). Similar to Table 5, except for the relatively low MAP of the LSTM-based method, the MAP values of the other five methods are all above 30%. It can be shown that our proposed AGS provides reliable data support for the theoretical research of SER in the recognition of environmental sound events. In addtion, from the practical results, the performance order (descending order) of these methods is: AST > Wavegram-Logmel-CNN > MobileNetV2 > Wavegram-Logmel-CNN-attention > DaiNet > LSTM-based Method. As shown in Table 6, the performance order (descending order) of these methods maintains the performance order in Table 5.

## 6    Conclusion and Future Work

In this paper, we propose a novel domestic scene sound dataset (AGS) for sound event recognition, which fills in the shortcomings of visual scene monitoring in the time dependence of events in environmental monitoring and the monitoring of the intensity and magnitude of the action behavior of the subject and object. Our work addresses the problem of incomplete data disclosure in Audioset due to policy permissions, and provides reliable data support for spatio-temporal-constrained activity monitoring scenarios. Meanwhile, we also compare state-of-the-art deep learning models to establish baseline SER performance on AGS. The detailed and sufficient experimental results can be shown that our proposed AGS provides reliable data support for the theoretical research of SER in the recognition of environmental sound events.

For future research work, we will continue to increase the amount of AGS data, and will consider exploring the generation of sound scene graphs based on AGS, and how to generate images or videos through environmental scene audio (focused on non-speech).